\documentclass{article}
\usepackage{setspace}
\usepackage{authblk}
\usepackage[T1]{fontenc}
\usepackage{breqn}
\usepackage{mathtools}
\usepackage{amsmath}
\begin{document}

\author[1]{Daniel P. Whitmire}

\title{The Abiogenesis Timescale}
\maketitle
Department of Mathematical Sciences,  University of Arkansas, Fayetteville

\begin{abstract}

If two physical timescales are independent, i.e. they depend on different physics, then (statistically) there is no reason to believe that their values should be equal, even to an order of magnitude. The timescale for abiogenesis $\tau_{AB}$, which depends primarily on prebiotic-chemistry, is expected to be independent of the planetary habitability timescale $\tau_{Hab}$, which depends primarily on the sun and therefore on nuclear forces and gravity. Therefore, we expect that either $\tau_{AB} \ll  \tau_{Hab}$ or $\tau_{AB} \gg \tau_{Hab}$.   The correct inequality is universal and a single example should suffice  to resolve this binary choice. Here I argue that, contrary to a well known anthropic selection effect,  our existence (which entails life on Earth) can be considered evidence that the correct choice is the former. A Bayesian analysis, taking into account that our existence is old evidence,  implies that the probability of the hypothesis $\tau_{AB} \ll  \tau_{Hab}$ is > 0.91, assuming equal priors. The Bayes factor, which depends only on the evidence, is > 10 and suggests strong to decisive support for the short abiogenesis timescale hypothesis, according to the Jeffreys interpretation.
\end{abstract}


\section{Background}

It has been argued that since we had to find ourselves on a planet where the origin of life occurred then nothing can be inferred about the probability of abiogenesis from this observation alone, other than it can't be zero. This anthropic selection effect is also used to invalidate the application of the principle of mediocrity to Earth, ${\it i.e.}$ Earth cannot be considered a typical Earth-like planet in the reference class of all such planets since Earth is not a random selection from this reference class.
In a previous paper (Whitmire 2023, hearafter paper I), I suggested that this logic is flawed because it doesn't take into account the fact that life on Earth is old/existing evidence. When this is taken into account, the likelihoods (probability of the evidence given the hypothesis) are evaluated counterfactually,  prior to abiogenesis. 

The old evidence issue for abiogenesis closely parallels the same issue as it applies to the multiverse hypothesis.
Namely, can our (old evidence) fine-tuned universe be considered as evidence for a multiverse?
After considering some of these arguments and conclusions, and citing clear examples where our existence can be evidence for an hypothesis, I return to the question of abiogenesis and whether our existence can be used as evidence in a Bayesian calculation of the probability of abiogenesis on planets similar to Earth. A crucial assumption in this analysis is a {\em statistical} independent-timescale argument which I discuss more fully here than in the earlier paper.

    
Carter (1983) gave an argument  which concluded that Life on Earth (LoE) cannot be considered evidence 
that abiogenesis on Earth-like planets is typical.  ${\it i.e.}$ the principle of mediocrity cannot be applied to Earth because of the selection effect of our existence.  Earth is not a random selection from the reference class of all Earth-like planets. Though not necessary, Carter presented his abiogenesis (AB) argument in a  binary Bayesian framework as follows (see also Paper I).  
Let the hypothesis to be tested = "AB is easy", and the compliment binary hypothesis = "AB is hard",  and let the evidence = "observation of LoE ".  Bayes' theorem for the posterior probability of the hypothesis is then 
    \begin{equation}
    \textnormal{P(AB easy|LoE)}=\frac
    {\textnormal{P(LoE|AB easy)P(AB easy)}}
    {\textnormal{P(LoE|AB easy)P(AB easy)+P(LoE|AB hard)P(AB hard)}}
    \end{equation}
    where P(AB easy) is the prior probability that AB 
    is easy in general before taking into 
    account the evidence of
    LoE and P(AB hard) is the prior probability 
    that AB is in general hard before taking into
    account that LoE exists. Binarity of hypotheses is 
    not a necessary assumption but sufficient to make the
    argument.  According to Carter, LoE exists regardless of 
    whether AB is easy or hard, consequently the 
    likelihoods 
    P(LoE|AB easy) = P(LoE|AB hard) = 1.
    Inserting this into Eq. (1) and noting 
    P(AB hard) + P(AB easy) = 1, gives
    \begin{equation}        
        P(\textnormal{AB easy}| \textnormal{LoE}) =  
        \frac{P(\textnormal{AB easy})}{P(\textnormal{AB easy})+
        P(\textnormal{AB hard})} = P(\textnormal{AB easy}),
    \end{equation}
    {\it i.e.}, the posterior probability that the hypothesis "AB is easy" 
    given the evidence LoE = prior 
    probability that AB is easy. 
    Therefore the observation of LoE is not
    evidence that AB is easy or hard, {\it i.e.}, the evidence of LoE does not update the 
    prior. Note that, if correct, this implies that in general once the evidence exists the likelihoods are always = 1 {\em independent} of the hypotheses.

In Paper I  it was pointed out that Carter's analysis is open to "The problem of old evidence" in Bayesian Confirmation Theory (Clymour 1980) and  three examples were given to illustrate the point.  One example,  similar to the one originally given by Clymour, is the flipping of a coin three times and getting three heads. The hypothesis is "the coin is fair". The analogous Carter-like approach would hold that the evidence already exists so the probability of the evidence = 1 regardless of whether the coin is fair or not fair and the likelihoods P(3heads|coin is fair)= P(3heads|coin is not fair) =  1 and therefore the prior is unaffected by the evidence. Recognizing that the evidence of 3 heads is old evidence, the correct approach is to evaluate the likelihoods contrafactually or historically, as recommended by Clymour, prior to the flipping of the coins. Assuming equal priors, and that the unfair coin weight for heads lies in the interval (0.5,1] the result for the posterior probability is P(fair|3heads) = [1/9, 1/2). 

In the second example, there are two urns,  A  and  B.  One urn contains 10 marbles labeled 1 - 10 and the other urn contains 1,000,000 marbles labeled 1 - 1,000,000. A fair coin is flipped and a marble is drawn from urn A.  The selected marble is labeled \#7. What is the probability that urn A contains 10 marbles? It could be said that the evidence of marble \#7 already exists and that therefore the probability of the evidence = 1 and  the two likelihoods P(\#7|urn A has 10 marbles) = P(\#7|urn B has 10 marbles) = 1 and therefore the posterior probability = prior probability = 1/2. The evidence of drawing \#7 did not update the prior probability of 1/2. However, this is clearly incorrect. Instead, the likelihoods should be P(\#7 prior to obtaining the evidence|urn A has the 10 marbles) = 1/10 and P(\#7 prior to obtaining the evidence|urn B has 1,000,000 marbles) = 1/1,000,000. The posterior probability P(urn A contains 10 marbles|\#7) = 0.999  (Bostrom 2002a and Paper I). This qualitatively intuitive result also follows from the Principle of Mediocrity since the marble \#7 is a typical marble number if it is drawn from the urn that contains 10 marbles but is highly atypical if drawn from the urn that contains 1,000,000 marbles.

The third and most relevant example considered in Paper I is the Conception Analogy since it closely parallels the actual AB problem of interest. The conception (origin) of you could have occurred if your parents used  
    contraception (conception hard = CH) or if they did not use
    contraception (conception easy = CE). 
    What is the probability of the hypothesis =
    CE, given the evidence that you exist?
    Bayes' formula for this posterior probability is
    \begin{equation}
        \textnormal{P(CE|you exist)} = 
        \frac{\textnormal{P(you exist|CE)P(CE)}}{\textnormal{P(you exist|CE)P(CE)} 
        + \textnormal{P(you exist|CH)P(CH)}}.              
    \end{equation}
    As in the Carter argument, it could be said that you exist 
    regardless of whether your conception was easy or hard so
    P(you exist|CE) = P(you exist|CH) = 1, and then 
    P(CE|you exist) = P(CE), or the posterior probability 
    = prior probability and thus the 
    evidence  "you exist" doesn't update the 
    prior. But, as in the case of the coin flips and the
    urn example, this is not
    the correct analysis. It's not the old evidence of 
    your existence 
    that's important but rather the prior-conditional
    likelihoods P(you {\it will} exist prior to your conception|CE) and 
    P(you {\it will} exist prior to your conception|CH). 
    For example, assume that without contraception the 
    mean probability of conception is  0.85 per year
    = P(you will exist prior to your conception|CE) and that with contraception the mean probability is 0.01 per year =
    P(you will exist prior to your conception|CH). Assuming equal priors, this gives the posterior 
    probability P(CE|you exist) = 0.99, which seems qualitatively intuitive. The assumption of equal priors is approximately valid in this case (https://www.statista.com/statistics/1248526/contraceptive-prevalence-any-methods-among-women/). 

Our treatment of old evidence in these examples and in the abiogenesis case of interest discussed below seems straight forward. 
In addition to these, history is replete with examples of old evidence being used and accepted as legitimate evidence to confirm theories.  The perihelion shift of Mercury was old evidence at the time Einstein published his General Theory of Relatively and Kepler's laws of planetary motion were well known when Newton published his law of gravity.  Much, if not all, of Darwin's evidence for his theory of evolution by natural selection was old evidence. Historically, old evidence was not considered a problem until Glamour (1980) claimed that old evidence was incompatible with Bayesian confirmation theory epistemology.  This may be a philosophical problem in general for Bayesian confirmation theory but, as a practical matter, in all cases considered here it is easily taken into account as the above examples illustrate.



In the case of AB we consider the hypothesis  "AB is easy in general on Earth-like planets", or more simply "AB easy",  and the evidence =  LoE.  Easy and hard abiogenesis are defined below and Earth-like is discussed briefly in Paper I. At present a number of rocky planets lying in their star's habitable zone (Kasting et al., 1993) have been discovered (https://www.space.com/s30172-six-most-earth-like-alien-planets.html, but none are yet known to actually have surface liquid water which is a requirement for an Earth-like potentially habitable planet. In arguments to follow I assume that such planets do exist. Taking into account that LoE is old evidence, we interpret the likelihoods to be the probability of the LoE evidence prior to AB.  Bayes' formula is then
    \begin{equation*}
        \textnormal{P(AB easy|LoE)} =  
\end{equation*} 
\begin{equation} 
\frac{\textnormal{P(LoE prior to AB|AB easy)P(AB easy)}}
        {\textnormal{P(LoE prior to AB|AB easy)P(AB easy)+P(LoE prior to AB|AB hard)P(AB hard)}}.
    \end{equation}
The likelihoods, whatever their true values may be, are no longer expected to be exactly = 1 as assumed in the Carter argument. Therefore, on the basis of old 
evidence considerations {\em alone}, that argument fails,  implying that LoE is not neutral evidence in support of the hypothesis that AB is easy. 
The fact that the likelihoods are evaluated prior to the occurrence of AB seemingly removes the anthropic selection effect that is used to invalidate the application of the Principle of Mediocrity. At this pre-biotic epoch, Earth can be considered a typical Earth-like planet in the reference class of pre-biotic Earth-like planets. Therefore, any conclusions regarding AB on Earth should apply to the reference class of all Earth-like planets. 

\section{Can our existence ever be used as evidence for anything?}

The Carter argument holds that if the evidence (our existence) is known then the likelihoods = 1, independent of the hypothesis. This is a particular example of the more general problem of old evidence that has been discussed extensively in the philosophy literature. The idea is so counter-intuitive that Glymour's (1980) original article discussing the problem is entitled "Why I am not a Bayesian". This controversy is fundamental in the argument for a multiverse. Can our old evidence of a fine-tuned universe be used as evidence for a multiverse? It could be claimed that the fact that our universe is fined tuned is old evidence and therefore cannot be considered evidence of a multiverse hypothesis in a Bayesian frame work (Leeds 2007). Others (correctly I believe) don't think this is a problem at all.  Juhl (2007) gives a fanciful example involving hostile aliens that shows that our existence can be used as evidence in support of an hypothesis.  Beyond that example, he notes other more mundane examples:  Our existence can be used as evidence against the hypothesis that an asteroid destroyed all life on Earth one million years ago; or, a pandemic led to the extinction of the human species 1,000 years ago, and so on. In footnote 2 Barnes (2018) further rebuts the Carter-like assumption that if the evidence is already known then the likelihoods =1.



  
\section{The Independent-Timescale Argument}

In Paper I an independent timescale argument was used to define easy and hard AB. If two timescales are independent (depend on different physics) then there is no reason to believe that they should be equal, even to an order of magnitude. Carter (1983) used this assumption in connection with his argument that intelligent life is rare. The two timescales in that case were the timescale for the evolution of intelligent life and the solar timescale. That argument is distinct from  Carter's AB argument. This timescale argument is statistical in nature. Of course it could occur that two independent timescales are nearly equal but in general this would be highly unexpected. For example, consider the timescale of Earth's orbital period, which depends on gravity, compared to the half-lives of all known radioactive nuclides, which depend on the strong and weak nuclear forces.  A table of approximately 3,000 radioactive nuclide half-lives can be found at https://en.wikipedia.org/wiki/List_of_radioactive_nuclides_by_half-life. The number of nuclides with half-lives that lie between 0.1 yr and 10 yr, {\it i.e.} within an order of magnitude of 1 yr, is 38. Therefore, only 38/3000 = 1.3\% of the nuclide half-lives are not consistent with the independent-timescale argument, leaving 98.7\% compliant, in this example. If we compare all half-lives with the 1.3 s rotational  period of the first discovered pulsar the result is 6\%, or 94\% compliant. If we compare all half-lives with the orbital period of the sun around the galaxy (230 Myr) the result is 0.5\% or 99.5\% compliant. The average of these three examples is 2.6\%, suggesting that the independent-timescale argument is typically $\sim$ 97\% accurate. 
The independent-timescale argument justifies reducing the hypotheses to only two, H1 = $\tau_{AB} \ll \tau_{Hab}$  and H2 = $\tau_{AB} \gg \tau_{Hab}$ (see below), which simplifies the analysis and interpretation of the results. 

\section{Abiogenesis}

 The timescale for abiogenesis on Earth-like planets $\tau_{AB}$  depends primarily on pre-biotic chemistry, while the planetary habitability timescale 
$\tau_{Hab}$ depends primarily on the sun (Kasting et al. 1993) and thus nuclear physics and gravity.  Therefore, from the independent-timescale argument, we expect that either $\tau_{AB} \ll \tau_{Hab}$  or $\tau_{AB} \gg \tau_{Hab}$.  We adopt these two inequalities as the natural definitions of easy and hard AB. These inequalities are {\em general} expectations for Earth-like planets orbiting solar type stars.  In Paper I, these unconditional inequalities were incorrectly associated with the unconditional prior probabilities. The prior probabilities of the inequalities are not the same  as the inequalities themselves. Instead, the inequalities should be associated with the "conditional priors" or likelihoods, giving
\pagebreak
\begin{equation*}
\textnormal{P(LoE prior to AB|AB easy)} = \textnormal{P(LoE prior to AB}|\tau_{\textnormal{AB}} \ll \tau_{\textnormal{Hab}}
) \sim 1
\end{equation*}
\begin{equation*}
\textnormal{P(LoE prior to AB|AB hard)} = \textnormal{P(LoE prior to AB}|\tau_{\textnormal{AB}} \gg \tau_{\textnormal{Hab}}) =\epsilon
\end{equation*}
where $\epsilon < 1/10$.  

Inserting these likelihoods into equation (4) gives 
\begin{equation}
\textnormal{P}(\tau_{\textnormal{AB}}\ll\tau_{\textnormal{Hab}}|\textnormal{LoE})  = \frac{1}{1+\epsilon\frac
	{\textnormal{P}\left(
		\tau_{\textnormal{AB}}\gg\tau_{\textnormal{Hab}}
	\right)}
	{\textnormal{P}\left(
		\tau_{\textnormal{AB}} \ll \tau_{\textnormal{Hab}}
	\right)}
}
\end{equation}
For equal priors the posterior probability of P($\tau_{AB}\ll\tau_{Hab}$|LoE) is $>$ 0.91.  


Another approach is to calculate the Bayes factor (BF) which takes into account only the evidence supporting the hypothesis (Jeffreys 1961) . For binary hypotheses, the BF is equal to the likelihood ratio, or odds, of the two hypotheses = 1/$\epsilon > 10$. According to Jeffreys' interpretation of the BF, a value greater than 10 implies that the evidence for the hypothesis is strong to decisive. 
To interpret the BF, consider the  extreme example of  a person flipping a coin 100 times, resulting in 100 heads. 
The hypothesis = "coin is fair".
Although mathematically that evidence could be undone with an astronomically large prior ($\geq 10^{30}$), in practice we don't need to know the prior (probability that the 
person flipping the coin is honest or a magician, etc) to conclude that the coin is not fair. Of course the case at hand here is not
extreme, but the point being that at a certain level of evidence the unknown priors are not likely to undo the weight of the evidence. The Jeffreys interpretation of the BF is one measure of that level.

Although other timescales could correctly be used in comparison with $\tau_{AB},  \tau_{Hab}$ seems most relevant. For example, we expect that the timescale for one Earth orbital period or the rotational period of the first pulsar should be >> or << $\tau_{AB}$. Through likely true, these comparisons are not interesting for obvious reasons. Note that for Earth, the fact that AB did occur in a time much less than the habitability time, this alone can't be used as evidence that in general  $\tau_{AB} <<  \tau_{Hab}$ due to the selection effect that our evolution required a significant amount of time and even if  $\tau_{AB} >>  \tau_{Hab}$, AB on Earth would still have to occur early (Spiegel D. \& Turner E., 2012).

Though the priors are unknown, there may be other arguments favouring  the hypothesis 
$\tau_{AB} \ll \tau_{Hab}$, such as an appeal to a multiverse. If there are a very large number of universes with different fundamental constants and/or laws we might expect to find ourselves in a universe that is relatively more life-friendly. This universe need not be the optimum life-friendly universe (ours apparently is not) but a universe that is a balance between life-friendly constants/laws and frequency/probability. Another argument is an appeal to the Self-Indication-Assumption (SIA) (Bostrom 2002b) which holds that your existence alone implies more observers in the universe since the more "slots" available the greater the prior probability that you would fill one.  Although SIA can explain the Doomsday Argument (Olum 2002),  it is currently considered controversial.

Finally, I consider a version of the Conception Analogy that more closely parallels the above abiogenesis analysis. To that end we ignore the known data for the prior probabilities of CE and CH, 0.85/yr and 0.01/yr, respectively. Let the hypothesis CE = "conception of humans is easy in general" and evidence = "you exist".   Bayes' formula for this posterior probability is
    \begin{equation*}
        \textnormal{P(CE|you exist)} = 
\end{equation*}
\begin{equation}
        \frac{\textnormal{P(you will exist prior to conception|CE)P(CE)}}{\textnormal{P(you will exist prior to conception|CE)P(CE)} 
        + \textnormal{P(you will exist prior to conception|CH)P(CH)}}.              
    \end{equation}
In analogy with the AB analysis, we define CE  as  "$\tau_c \ll \tau_f $" and CH as 
"$\tau_c \gg \tau_f $", where $\tau_c$ is the conception timescale and $\tau_f$ is the fertility timescale. Then P(you will exist prior to conception|CE) $\sim$ 1,  and setting P(you will exist prior to conception|CH) = $\epsilon < 1/10$, and assuming equal priors, gives for the posterior probability P(CE|you exist) $> 0.91$. 

For the analogy to be complete it must be assumed that you are in an epistemic state of ignorance about $\tau_c$ and $\tau_f$ and do not know that they are not independent and that $\tau_c \ll \tau_f $. You know that might be true but it could also be that $\tau_c \gg \tau_f $ and you just got lucky. The posterior probability  P(CE|you exist) applies to you, but you are aware of the existence of other pre-conception couples.  Since the likelihood probabilities in our Bayesian analysis are determined pre-conception, we can extend the ressult to other pre-conception couples. By analogy with our (assumed)  knowledge of other pre-biotic Earth-like planets,  the likelihoods in that Bayesian analysis are determined in the pre-biotic state,  and thus those results should be extendable to other pre-biotic Earth-like planets. 


\section{Conclusion}

The recognition that life on Earth is old evidence and must be treated accordingly in a Bayesian framework is sufficient to challenge the Carter abiogenesis argument, which concludes that this evidence is neutral and does not support the hypothesis that abiogenesis is probable on Earth-like planets. In the case of old evidence such as life on Earth, the likelihoods should be evaluated prior to the occurrence of the evidence.The fact that the likelihoods should be evaluated prior to the occurrence of abiogenesis on Earth removes the anthropic selection effect that would otherwise invalidate the application of the principle of mediocrity to Earth. Prior to abiogenesis, Earth is  no different than any other pre-biotic Earth-like planet and so can be considered a typical member of this reference class.
Although we had to find ourselves on a planet where abiogenesis occurred, that observation alone says nothing about the probability of the process itself, other than it can't be zero.  A statistical independent-timescale argument implies that the timescale for abiogenesis on Earth-like planets should be either 
$\ll \tau_{Hab}$ or $\gg \tau_{Hab}$.  On the basis of the old evidence (life on Earth) alone, I find that the former is the more probable conclusion. 
Our analysis shows that, with the assumption of equal priors, the probability that the timescale for abiogenesis  is $\ll \tau_{Hab}$ is > 0.91. In terms of the Bayes factor (> 10), as interpreted by Jeffreys, the evidence for the hypothesis $\tau_{AB}\ll \tau_{Hab}$  is strong to decisive.

   \section*{ACKNOWLEDGMENTS} 

I thank 
Robert Graham for a critical review of an earlier version of this paper.

\section*{DATA AVAILABILITY} The radioactive nuclide half-life data used in the independent- timescale argument was obtained from 
https://en.wikipedia.org/wiki/List_of_radioactive_nuclides_by_half-life updated November 25, 2024. 

{}
\end{document}